\documentclass[aps,twocolumn,showkeys]{revtex4}

\usepackage[latin1]{inputenc}
\usepackage{graphicx}
\usepackage{epstopdf}
\usepackage{dcolumn}
\usepackage{bm}
\usepackage{amsmath}

\begin{document}


\title{Cavity-enhanced optical detection of carbon nanotube Brownian motion} 



\author{S. Stapfner}
\affiliation{Center for NanoScience and Fakult\"{a}t f\"{u}r Physik, Ludwig-Maximilians-Universit\"{a}t M\"{u}nchen, Geschwister-Scholl-Platz 1, 80539 M\"{u}nchen, Germany}

\author{L. Ost}
\affiliation{Center for NanoScience and Fakult\"{a}t f\"{u}r Physik, Ludwig-Maximilians-Universit\"{a}t M\"{u}nchen, Geschwister-Scholl-Platz 1, 80539 M\"{u}nchen, Germany}

\author{D. Hunger}
\affiliation{Center for NanoScience and Fakult\"{a}t f\"{u}r Physik, Ludwig-Maximilians-Universit\"{a}t M\"{u}nchen, Geschwister-Scholl-Platz 1, 80539 M\"{u}nchen, Germany}

\author{J. Reichel}
\affiliation{Laboratoire Kastler Brossel, Ecole Normale Sup\'{e}rieure, Universit\'{e} Pierre et Marie Curie, CNRS, 24 rue Lhomond, 75005 Paris, France}

\author{I. Favero}
\affiliation{Laboratoire Mat\'{e}riaux et Ph\'{e}nom\`{e}nes Quantiques, Universit\'{e} Paris-Diderot, Sorbonne Paris Cit\'{e}, CNRS, UMR 7162, 10 rue Alice Domon et L\'{e}onie Duquet, 75013 Paris, France}

\author{E.\,M. Weig}
\email[]{correspondance: weig@lmu.de}
\affiliation{Center for NanoScience and Fakult\"{a}t f\"{u}r Physik, Ludwig-Maximilians-Universit\"{a}t M\"{u}nchen, Geschwister-Scholl-Platz 1, 80539 M\"{u}nchen, Germany}


\date{\today}

\begin{abstract}
Optical cavities with small mode volume are well-suited to detect the vibration of sub-wavelength sized objects. Here we employ a fiber-based, high-finesse optical microcavity to detect the Brownian motion of a freely suspended carbon nanotube at room temperature under vacuum. The optical detection resolves deflections of the oscillating tube down to $\rm50\,pm/Hz^{1/2}$. A full vibrational spectrum of the carbon nanotube is obtained and confirmed by characterization of the same device in a scanning electron microscope. Our work successfully extends the principles of high-sensitivity optomechanical detection to molecular scale nanomechanical systems.
\end{abstract}



\maketitle 



%
%

%

Probing the vibrational motion of nano-scale objects has great potential for advancing next-generation technologies such as resonant mass or bio-sensing \cite{jensen_natnanotech2008,chaste_natnanotech2012,hanay_natnanotech2012}. A key example is carbon nanotubes (CNTs), which promise to display ultimate sensitivities due to their molecular scale mass and diameter along with their outstanding mechanical properties. Recently CNTs have attracted attention for their ability to resolve and exploit the quantum nature of mechanical vibration in optomechanical experiments \cite{wilson-rae_NJP2012,schneider_arxiv2012}. However, small vibrational amplitudes and dimensions impose severe challenges in the realization of CNT-based mechanical devices. One fundamental challenge of particular interest is to resolve the thermally excited Brownian motion of CNTs. Indeed this level of resolution allows operation of the undriven device in the linear regime with enough dynamic range, and limits the influence of spurious non-linear effects observed in CNTs \cite{eichler_nanolett2011,eichler_nature2011}. Transmission and scanning electron microscopy (TEM, SEM) have both been employed successfully to visualize the thermal motion of CNTs\cite{treacy_nature1996, babic_nanolett2003}, yielding a superposition of the small-amplitude envelopes of all oscillating modes. Yet electron beam imaging is a destructive method for observing CNT oscillation because of the unavoidable deposition of amorphous carbon on the tube. Electrical schemes provide a powerful approach for detection of vibrating CNTs, but typically require a coherent drive voltage to actuate the tube\cite{huettel_nanolett2009,steele_science2009,chaste_apl2011}. Furthermore their sensitivity is not sufficient to resolve the Brownian motion. Optical techniques are ultra-sensitive\cite{arcizet_2006prl} but the optical detection of CNT motion is hindered by the diffraction limit, with typical tube diameters much smaller than the wavelength of visible light. Dark field illumination was combined with confocal microscopy to detect the mechanical oscillation of driven CNTs with a diameter of $\rm80\,nm$, however, the Brownian motion amplitude could not be resolved with this approach \cite{fukami_JJApplPhys2009}. Even in the subwavelength regime, the sensitivity can be enhanced by using an optical cavity\cite{favero_2008}. In earlier work\cite{favero_optexp2009} we apply a fiber-based optical micro-cavity of small mode volume and high finesse to measure the Brownian motion of an amorphous carbon based nanorod with a diameter of about $\rm100\,nm$. Following the proposal of extending cavity optomechanics experiments to CNTs\cite{favero_2008}, in the present work the nanorod is replaced by a carbon nanotube with a ten times smaller diameter. Taking advantage of an improved cavity with significantly increased finesse, we present data clearly resolving the Brownian motion of the CNT oscillating within the cavity light field. Research in optomechanics \cite{kippenberg_science2008,marquardt_physics2009, favero_natphot2009, aspelmeyer_joptsoc2010} will benefit from the in-cavity implementation of carbon nanotubes, which are the smallest solid-state mechanical resonators available to date, and have not yet been explored in this context.\\

\begin{figure*}[htb]
\includegraphics{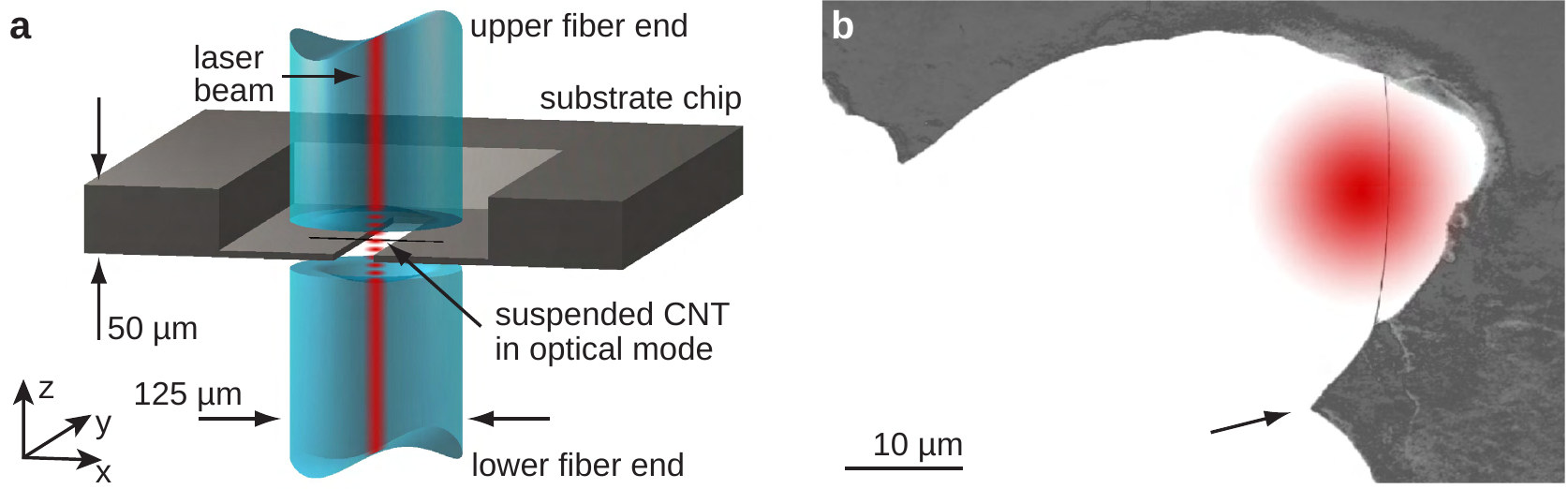}%
\caption{\label{cavity-cnt}(COLOR ONLINE) Illustration of the experimental setup. (a) Schematic view of the cavity and sample chip with the CNT introduced into the cavity mode. (b) Illustration based on an SEM image of the CNT under investigation suspended across the gap (white) with surrounding substrate (grey). The red spot illustrates the position and extension of the optical cavity mode during the experiment.}%
\end{figure*}

Without exploiting an optical transition, direct optical detection of a CNT of deep subwavelength dimensions is a challenging task\cite{sawano_nanolett2010}. In an optical cavity, however, the dispersive and dissipative interactions between a nanotube and the light field is enhanced. Particularly for the case of a high finesse cavity of small mode volume, this gives rise to an amplification of the signature of the CNT displacement in the cavity response. For the case of a dispersive interaction, the CNT imprints a phase shift on the photons in the cavity, which results in a resonance frequency shift of the cavity mode. On the other hand, a dissipative interaction, for example caused by absorption and scattering of photons by the nanotube, leads to a net loss of energy decreasing the circulating optical power. The strength of both types of interaction depends on the position and orientation of the nanotube within the cavity mode. Thus by measuring the frequency and intensity of the cavity resonant response, information can be gained on these interactions. In the present work both dispersive and dissipative interactions of the CNT contribute to the observed signal.

To detect the CNT's Brownian motion, we employ a fiber-based micro cavity formed by two opposing, concavely shaped fiber end facets, which are coated with a highly reflective Bragg mirror optimized for the cavity wavelength of $\rm780\,nm$\cite{steinmetz_2006,hunger_njp2010, hunger_AIP2012, flowers_arxiv2012}. Fig. \ref{cavity-cnt}a displays a schematic of the setup. For a mirror separation of $\rm37\,\mu m$, the cavity optical mode waist radius is $\rm3.4\,\mu m$ and the finesse is measured to be $\rm24,500$. Alignment presents a major technical challenge of this configuration in conducting optomechanical experiments, as illustrated by Fig.\ref{cavity-cnt}a. In order to enter a nanomechanical object into the cavity, the hosting substrate chip must fit into the narrow slit formed between the two fibers. This implies that the substrate thickness has to be reduced below the width of that slit for an area covering the fiber cross section corresponding to their diameter of $\rm125\,\mu m$. The center of this thinned out area hosts a gap wide enough to avoid clipping losses from its edges, which would disrupt the optical mode. According to numerical simulations, the gap has to be at least $\rm18\,\mu m$ wide to fulfill this condition. For ease of alignment with the wavefronts of the cavity mode the CNT is doubly clamped and freely suspended across this gap as indicated in Fig. \ref{cavity-cnt}a.

Suitable substrates serving as holders for the CNTs are fabricated from a $\rm50\,\mu m$ thick silicon wafer using optical lithography and wet etching processes from both the top and the back side of the wafer (see supplement). Subsequently CNTs are grown across the gap using a process adapted from Ref. \onlinecite{babic_nanolett2003}. Careful SEM and TEM inspection of different samples reveal that the doubly-clamped CNTs are freely suspended across gaps of up to $\rm25\,\mu m$. Additional electron diffraction analysis in the TEM show that the resulting CNTs are multi-walled tubes or thin ropes consisting of $\rm5$ to $\rm7$ individual tubes. Details of chip fabrication, growth and SEM/TEM analysis are provided in the supplement. Figure \ref{cavity-cnt}b shows a post-processed SEM image of the suspended CNT investigated in the present work. The suspended section is $\rm19.7\,\mu m$ long with a diameter between $\rm6$ and $\rm8\,nm$. The slightly wavy shape of the tube indicates the absence of tensile stress along the tube. Close-up micrographs (see Fig. \ref{cntexcitation}) reveal a feature in the center of the CNT, which might be catalyst residue from the growth process\cite{song_JPCC2008}.

\begin{figure}[tbh]
\includegraphics{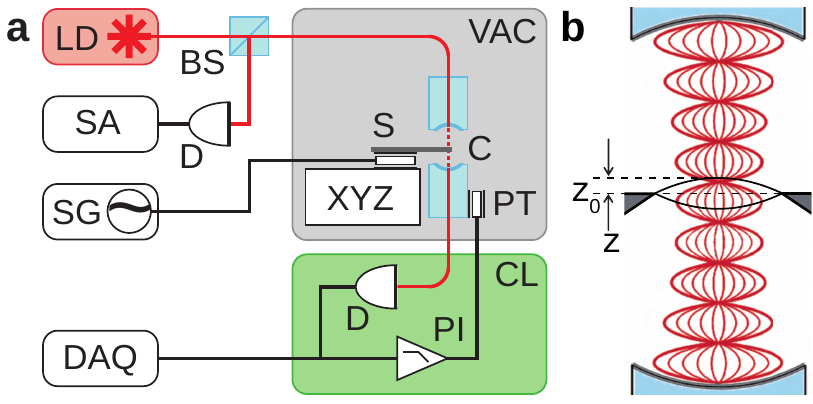}%
\caption{\label{setup-cavity}(COLOR ONLINE) (a) Schematics of the experimental setup. (b) Illustration of the cavity with the CNT resonating with amplitude $z$ around equilibrium position $z_0$ in the optical mode.}%
\end{figure}

\begin{figure*}[hbt]
\includegraphics{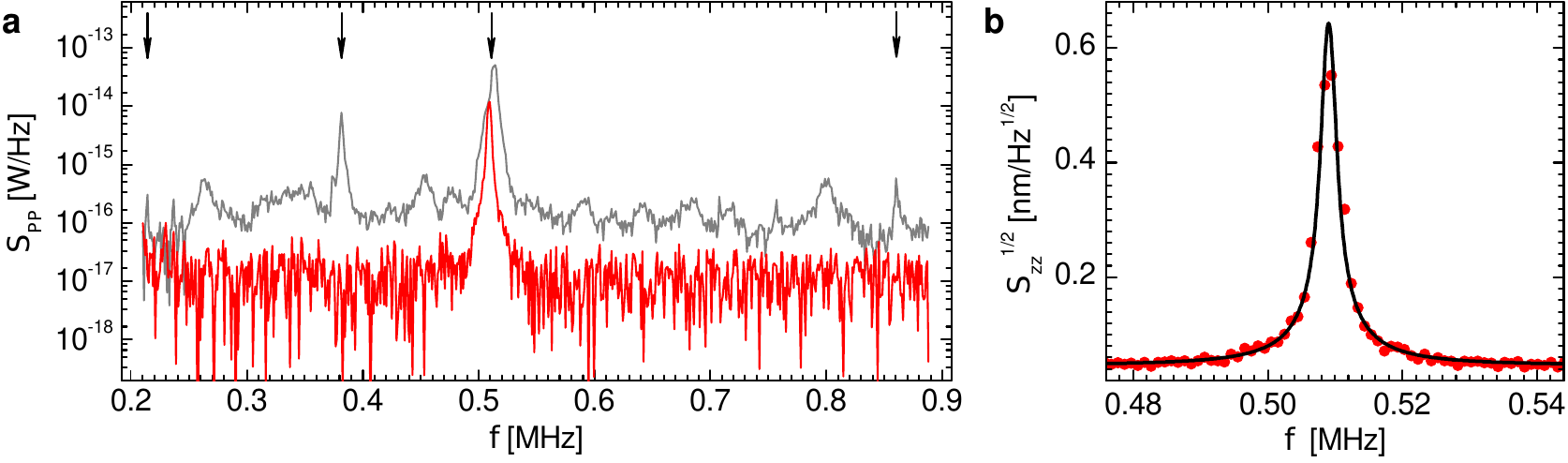}%
\caption{\label{dataplots}(COLOR ONLINE) (a) Power spectra with subtracted background reveal the CNT's vibrational spectrum. The gray trace is obtained with the CNT placed in the cavity mode and $\rm14\,dBm$ of white noise is applied to the drive piezo, whereas for the red trace the drive is switched off. The remaining peak near $\rm0.5\,MHz$ rising more than $\rm20\,dB$ above the noise floor evidences the Brownian motion of the CNT. Black arrows point at the peaks confirmed as mechanical resonances of the CNT by subsequent SEM experiments. (b) Zoom of the CNT's Brownian noise peak with Lorentzian fit (black) and data points calibrated to Brownian vibrational amplitudes.}
\end{figure*}

For optical measurements of CNT vibration the substrate chip is introduced into the fiber-based cavity as illustrated in Fig. \ref{cavity-cnt}a (a photograph of this experimental situation is shown in the supplement). In order to circumvent gas damping effects\cite{fukami_JJApplPhys2009} on the mechanical motion, the experiments are carried-out in a custom fiber-compliant vacuum cell (VAC) at a pressure of $\rm10^{-5}\,mbar$ and at room temperature ($\rm300\,K$), see Fig. \ref{setup-cavity}a. A three axis XYZ-positioner allows accurate placement of the CNT inside the cavity (C)(see Fig. \ref{setup-cavity}b). Optionally a piezo transducer (PT) underneath the sample chip (S) can be used to excite the CNT's mechanical motion via a signal generator (SG). The cavity is pumped with a stabilized diode laser (LD) at $\rm780\,nm$. The transmitted light is sent to a photo detector (D), monitored on an oscilloscope (DAQ) and used to lock the cavity (CL) on a slope of an optical resonance. Therefore an electronic feed-back loop (PI) acts on a piezo (PT) controlling the cavity length. At the beam splitter (BS) the light reflected from the cavity is directed to a second photo detector (D). The CNT vibration is measured by analyzing the optical noise of the reflected light (spectrum analyzer SA). More details on the cavity vacuum setup as well as laser and cavity stabilization schemes can be found elsewhere\cite{stapfner_spie2010}. In order to be sensitive to both the dissipative and dispersive component of the CNT-light interaction, the cavity was locked on the slope of its optical resonance. The CNT displacement is probed by monitoring changes in the reflected and transmitted light intensities which are read out by photodetectors (D).

The optical power transmitted on the empty cavity resonance is $\rm0.56\,\mu W$. Bringing the CNT into the optical mode, as illustrated by the red spot in Fig. \ref{cavity-cnt}b the cavity transmission drops by $29\,\%$. This drop is predominantly caused by residual clipping losses originating from the presence of the hosting substrate edges near the CNT. When locked on the cavity resonance slope, the vibrational motion of the CNT is characterized under external actuation from the drive piezo. Subject to excitation with white noise ($\rm14\,dBm$ with $\rm20\,MHz$ bandwidth), a series of resonances is clearly observable above the noise floor (Fig. \ref{dataplots}a gray trace). Spectra shown in Figure \ref{dataplots}a are obtained after substraction of the reference background of the empty cavity. The main peak at $\rm0.51\,MHz$ is the first flexural mode of the tube. This peak and other spectral features (black arrrows) have been identified as mechanical resonances of the CNT through SEM experiments complemented by beam theory calculations and signal amplitude estimations, as we will detail below. This main peak at $\rm0.51\,MHz$ rises up from the noise floor by more than $\rm20\,dB$, even if the piezo is not driven (Fig. \ref{dataplots}a red data). A Lorentzian fit to the data points of the undriven resonance (Fig. \ref{dataplots}b red line) yields a quality factor of $\rm300$. Interestingly, this signal peak is not stable in frequency but fluctuates slowly between $\rm0.47\,MHz$ and $\rm0.52\,MHz$ on the timescale of minutes. We suggest slack and conformational instability of the tube, which are also observed in the SEM, contribute most to this behavior. Furthermore, a strong dependence of the signal amplitude on the position of the CNT in the optical mode is observed. The resonance essentially vanishes when the CNT is positioned a few micrometers away from the optical mode axis. Other information can be gained when the sample is positioned such that the optical mode lies on a protrusion of the hosting substrate (for example at the position indicated by a black arrow in Fig. \ref{cavity-cnt}b, about ten micrometers away from the CNT). At this position, with the protrusion entering the optical mode, the drop in cavity transmission is still of order $\rm29\,\%$ but the spectrum exhibits a series of extra peaks below $\rm0.22\,MHz$ most probably orginating from vibrating substrate modes (see supplemtental material). Such substrate resonances are not observed above $\rm0.22\,MHz$, indicating that the main resonance at $\rm0.51\,MHz$ does not stem from a substrate mode. The cavity resonant transmission exhibits constant values for the CNT being placed at several $z$-positions inside the cavity. This indicates that the dissipative signal component is dominated by clipping losses such that the CNT presence does not contribute to total extra cavity loss in our experiment. This implies that the nanomechanical motion detection relies primarily on the dispersive interaction.

To obtain an estimate for the resonance frequency $f_0=\omega_0/(2\pi)$ of the fundamental flexural oscillation of the measured CNT, conventional Euler-Bernoulli beam theory is applied, which models the mechanical properties of CNTs satisfactorily\cite{babic_nanolett2003,garcia_prl2007,martin_apl2007}. The model of an unstressed beam\cite{timoshenko} leads to

\begin{equation}
  \omega_0 = \frac{\beta^2}{l^2}\sqrt{\frac{Y\,I}{\rho\,A}},
\label{f_res}
\end{equation}

where $\beta = 4.72$ for the fundamental mode and $l = \rm19.7\,\mu m$ is the length of the nanotube. Based on our TEM analysis we consider the observed CNT to be either a multiwall nanotube or a rope comprising five to seven individual tubes with a total outer diameter $d_\text{o}$ between $\rm6$ and $\rm8\,nm$. The area moment of inertia $I = \pi(d_\text{o}^4-d_\text{i}^4)/64$ and the cross-section area $A = \pi (d_\text{o}^2-d_\text{i}^2)/4$ are used for multiwall nanotubes with the inner diameter $d_\text{i} = d_\text{o}-n\,\Delta_\text{s}$, the number of walls $n$ and the inter wall spacing $\Delta_\text{s} = \rm0.335\,nm$, and for ropes with $d_\text{i} \rm = 1\,nm$. The physical mass density $\rho$ is adopted from the mass density of graphite $\rho_{\text{graphite}} = \rm 2200\,kg/m^3$ along with $n$, $\Delta_\text{s}$ and by geometrical considerations for both cases. Values for the Young's modulus $Y$ in the range between  $\rm0.3$ and $\rm1\,TPa$ can be found in literature\cite{salvetat_advmater1999, salvetat_prl1999, ruoff_crphys2003,loeffler_ultramic2010}. Entering the above input parameters into Eq. \ref{f_res} yields an estimated resonance frequency between $\rm0.16$ and $\rm0.63\,MHz$, which is consistent with the measured frequency of the CNT.

The equipartition theorem allows calibration of the cavity noise peak in Fig. \ref{dataplots}a to the Brownian motional amplitude of the CNT using

\begin{equation}
  \frac{1}{2}k_B\,T = \frac{1}{2}k\left\langle z^2\right\rangle,
\label{equipartition}
\end{equation}

with the Boltzmann constant $k_B$, room temperature $T = \rm300\,K$, spring constant $k = \omega_0^2\,m_{\text{eff}}$. The mean squared amplitude $\left\langle z^2\right\rangle$ is the integral of the position squared spectral density in frequency space $\int^{\infty}_{0}S_{zz}(\omega)d\omega$ and the effective mass $m_{\text{eff}} = 0.81\cdot m = \rm0.3\dots1.7\,fg$ for the fundamental mode\cite{ekinci_2005}(see supplement) is derived from the physical mass $m = \rho\,l\,A$. The large uncertainty in the mass comes from the above-mentioned uncertainty in the composition and diameter of the tube and translates into an error of $\rm50\,\%$ for the calibration shown in Fig. \ref{dataplots}b. The nanotube's Brownian motion peak amplitude of $\rm0.6\,nm\pm50\,\%$ was measured with a sensitivity of $\rm50\,pm/Hz^{1/2}\pm50\,\%$.

Assuming a purely dispersive optomechanical interaction, the effect of this displacement of the CNT on the cavity reflection can be estimated using the dispersive cavity frequency shift\cite{chang_procnas2010}

\begin{equation}
  \frac{\delta\omega_c}{\omega_c} = -\frac{1}{2}\frac{\int\delta\mathbf{P}(\mathbf{r})\cdot\mathbf{E}(\mathbf{r})\,d^3\mathbf{r}}{\int\epsilon_0\,\mathbf{E}^2(\mathbf{r})\,d^3\mathbf{r}},
\label{frequencyshift}
\end{equation}

known from cavity perturbation theory\cite{waldron_prociee1960}, with the electric field amplitude $\mathbf{E}(\mathbf{r})$, polarization $\delta\mathbf{P}(\mathbf{r})=\epsilon_0(\epsilon_r-1)\mathbf{E}(\mathbf{r})$, dielectric constant $\epsilon_0$ and permittivity for CNTs\cite{heer_science1995} $\epsilon_r \approx 3$. For the first flexural mode of the CNT and in case of perfect alignment of the tube orthogonal to the $\rm TEM_{00}$ cavity mode axis, the integral of Eq. \ref{frequencyshift} leads, for our numerical parameters, to a maximal frequency-pull coupling rate $g_\text{om}=\partial\omega_c/\partial z$ of $\rm(2\pi)\,1\,MHz/nm$ along the cavity axis. With the employed laser power, detector response and optimized dispersive detection by placing the CNT in the middle between a node and an antinode of the optical field, the cavity reflection noise power should peak at $\rm1\,pW/Hz$ for the Brownian motion of the CNT. To explain the observed peak value of $\rm10\,fW$, we have to consider the CNT being tilted with respect to the wavefronts of the cavity field by an angle of $\rm 7^{\circ}$, which is compatible with the tolerance of alignment by binocular inspection. This independently supports the identification of the $\rm0.51\,MHz$ resonance as Brownian motion of the first flexural mode. 

\begin{figure}[bth]
\includegraphics{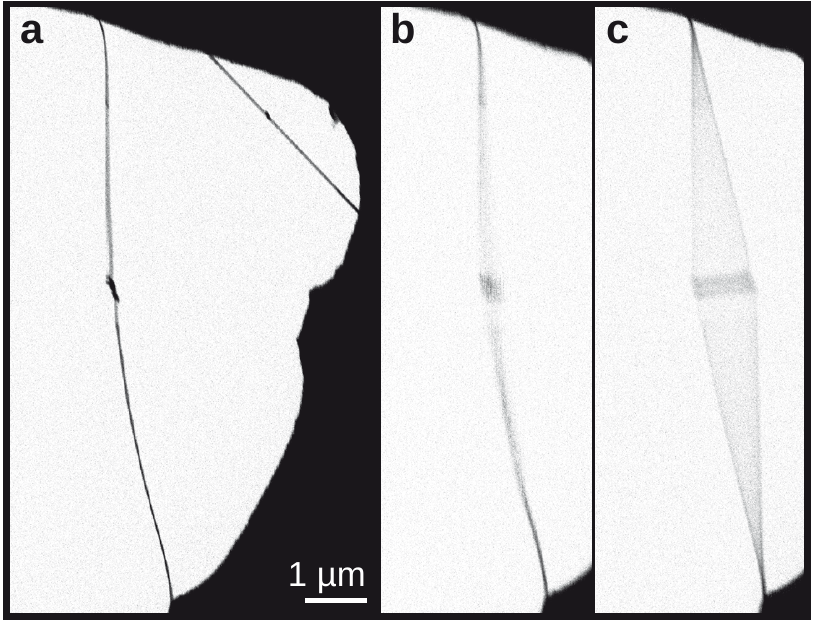}%
\caption{\label{cntexcitation}SEM images showing the CNT (a) at rest and excited through the drive piezo with $\rm14\,dBm$ (b) of white noise and (c) at $\rm518\,kHz$. Colors are inverted such that the substrate and CNT appear black and the gap is light gray.}
\end{figure}

To verify the nature of the observed signal, we carry out in-depth SEM investigation on the suspended CNT sample. This investigation starts after completion of the optical measurements in the cavity, in order to avoid electron beam induced contamination. As shown in Fig. \ref{cntexcitation}a, the CNT is still present at its original position and appears visually unchanged after being exposed to the intense light field of the cavity of about $\rm24\,kW\,cm^{-2}$. In order to visualize the CNT vibrational modes in the SEM, external actuation through the drive piezo is employed. Using the same white noise signal as applied in Fig. \ref{dataplots}a the CNT  image blurred, showing the envelope of its flexural oscillation modes (see Fig. \ref{cntexcitation}b) with a large deflection in the center\cite{babic_nanolett2003}. Note that when driven with a broad band source, the nanotube oscillates on several modes simultaneously such that the envelope consists of the superposition of the envelopes of all excited modes. Driving the sample with a sinusoidal signal whose frequency is swept isolates individual vibrational modes in the range between $\rm0.1$ and $\rm1\,MHz$. This allows us to identify all peaks from the optical cavity spectrum marked by black arrows in Fig. \ref{dataplots}a and in supplemental material as CNT mechanical modes at $\rm153\,kHz$, $\rm169\,kHz$, $\rm214\,kHz$, $\rm381\,kHz$, $\rm518\,kHz$, $\rm860\,kHz$, $\rm904\,kHz$ and $\rm917\,kHz$. Figure \ref{cntexcitation}c shows the CNT oscillating of the first flexural mode when driven by a $\rm14\,dBm$ power sinusoidal signal at $\rm518\,kHz$. Measuring the vibrational amplitude and sweeping the drive frequency across this resonance at much lower powers allows to estimate the quality factor to be about $\rm250\pm50$ which comes close to the value measured in the cavity.
It has to be noted that electron beam induced deposition on the CNT affects the mechanical properties, but by exposing the tube to the electron beam only at small areas during frequency scanning and taking single shot images this effect can be minimized such that during the measurements no obvious down-shift in resonance frequency was observed.

One possible effect distorting the calibration of the CNT Brownian amplitude is heating of the CNT by absorption of laser light. Neglecting optical losses by clipping on the hosting substrate and scattering losses induced by the CNT, we envision the extreme case where all optical power lost by the cavity is absorbed by the CNT and turns into heat. Using a  heat transfer model for the heat flow $\dot{Q}=\kappa\,A\,\Delta T/l$ with the thermal conductivity\cite{hsu_nanolett2009} $\kappa = \rm118\,W/(m\,K)$ of a comparable CNT of length $l$ and cross-section $A$ yields an increase in temperature $\Delta T$ of $\rm150\,K$. Under the above assumptions the CNT would oscillate at an elevated vibrational temperature with larger Brownian amplitude and we would have to correct the measurement calibration by a factor $\rm1.2$, which is smaller than the uncertainty coming from the CNT mass estimation.

Furthermore optomechanical back-action could contribute to actuate the CNT and modify the mechanical resonance and linewidth\cite{favero_2008}. Using a formalism to analyze optomechanical effects\cite{metzger_2008} in our data, we estimate that the change in effective vibrational temperature due to dynamical back-action would be at most of a few kelvins, hence negligible for our present study. Further, we extract similar Q-values from measurements in the cavity and in the SEM, strongly indicating that the motion is not driven optomechanically during measurements.

The relatively large zero point amplitude of CNTs $z_{\text{zpf}} = (\hbar/(2\,m\,\omega_0))^{1/2}$ makes them an interesting candidate for optomechanical studies. For the tube investigated in the present article $z_\text{zpf}\rm=5.2\,pm$. With an improved yet technically feasible cavity with line width $\kappa_c/(2\,\pi) \rm = 20\,MHz$ and for good alignment of the CNT to the optical mode, the zero point equivalent optomechanical coupling rate $g_0 = g_\text{om}\,z_{\text{zpf}}\rm = (2\pi)\,33\,kHz$ is expected. A cryogenic environment usually boosts the CNT's mechanical quality factor\cite{steele_science2009} beyond $10^{5}$, which would place the system deeply into the strong coupling regime\cite{teufel_nature2011a,safavi-naeini_nature2011,weis_science2010}. This is apparent form the cooperativity $C = 4g_0^2\,n_c\,Q/(\kappa_c\,\omega_0)$, an important figure of merit for the application of optomechanical systems, which relates the zero point optomechanical coupling rate $g_0$, the intracavity photon number $n_c$, the cavity linewidth $\kappa_c$ and the mechanical damping rate $Q/\omega_0$. For the described parameters a cooperativity per photon $C_0 = C/n_c = 1.1$ can be expected, which is higher than that reported in state-of-the-art optomechanical experiments\cite{teufel_nature2011a,safavi-naeini_nature2011,weis_science2010}.\\


In summary, we developed thinned silicon substrates with freely suspended carbon nanotubes suitable for measurements in fiber-based optical micro cavities. Taking advantage of the small mode volume of such a cavity we clearly resolve the Brownian motion of a CNT, a mechanical nanostructure much smaller than the wavelength of the employed light, in the optical cavity reflection. Optically measured resonance frequencies are confirmed by measurements in the scanning electron microscope. In contrast to SEM imaging techniques, the optical detection technique presented here allows much higher motional sensitivity, integration in a device based on fiber optics, and does not contaminate the CNT. Furthermore we anticipate these advances to lead to novel optics based carbon nanotube architectures that will allow probing of the quantum nature of molecular mechanical systems and improvement in their performance.

\begin{acknowledgments}

We would like to thank Prof. Sch\"{o}nenberger and Markus Weiss who kindly introduced us into the art of growing clean carbon nanotubes, Dr. D\"{o}blinger for TEM imaging and analysis of the CNTs, Prof. Kotthaus and Prof. Karra\"i for fruitful discussions and  Darren R. Southworth for critically reading the manuscript. We gratefully acknowledge finacial support from the German-Israeli Foundation (GIF), the German Excellence Initiative via the Nanosystems Initiative Munich (NIM), the German and French Academic Exchange Service (DAAD and EGIDE Procope program) and the Bayerisch-Franz\"{o}sisches Hochschulzentrum (BFHZ).

\end{acknowledgments}

\cleardoublepage

\setcounter{figure}{0}
\setcounter{equation}{0}
\setcounter{table}{0}
\setcounter{page}{1}

\newcommand*{\mycommand}[1]{\texttt{\emph{#1}}}
\renewcommand{\figurename}{Supplementary Figure}
\renewcommand{\tablename}{Supplementary Table}
\renewcommand{\thefigure}{S\arabic{figure}}
\renewcommand{\thetable}{S\arabic{table}}
\renewcommand{\theequation}{S\arabic{equation}}
\newcommand{\fig}[1]{Supplementary Fig.~\ref{#1}}
\newcommand{\tab}[1]{Supplementary Tab.~\ref{#1}}
\newcommand{\eq}[1]{Eq.~(\ref{#1})}

\onecolumngrid

\part*{Supplementary Information: Cavity-enhanced optical detection of carbon nanotube Brownian motion} 

\section{Sample fabrication and characterization}

\subsection{Substrate fabrication}

\begin{figure*}[hb]
\includegraphics{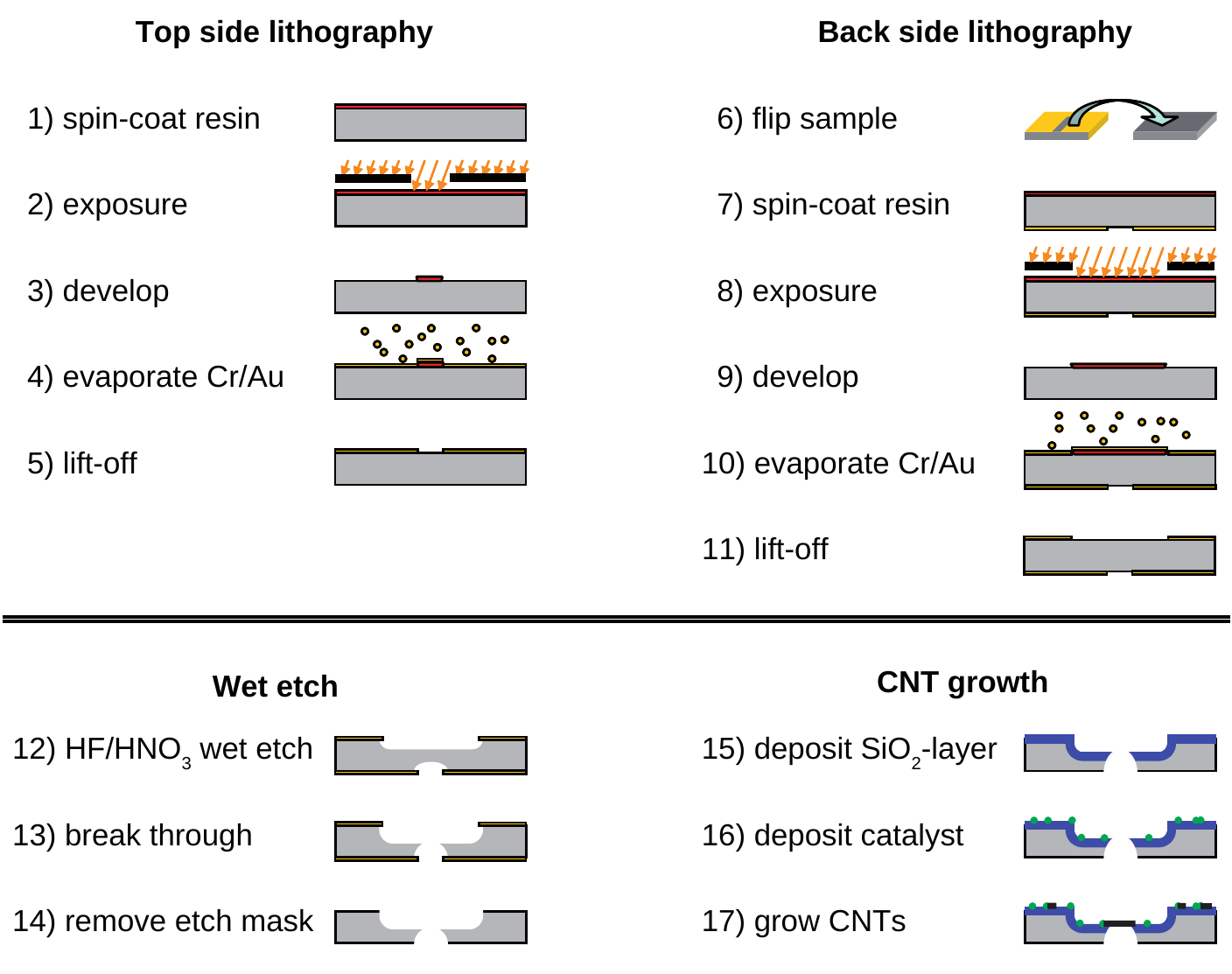}%
\caption{\label{samplefab} Sample processing step by step.}%
\end{figure*}

Sample fabrication processing steps are depicted in \fig{samplefab}. Suitable substrate chips to serve as holders for the CNTs were fabricated starting from a $\rm50\,\mu m$ thick silicon wafer. Photo lithography and a lift-off process after evaporation of $\rm4\,nm$ Cr and $\rm50\,nm$ Au (steps 1 through 11) performed on both sides of the wafer define the etch mask for the subsequent isotropic silicon wet etch, ($\rm HNO_3\,(65\,\%) : HF\,(49\,\%) = 98 : 2$) for about $\rm60\,min$ (setps 12 and 13). After the etch mask was removed (step 14) in aqua regia ($\rm HCl\,(37\,\%) : HNO_3\,(65\,\%) = 3 : 1$) the chips were rinsed with water and isopropanol and dried with a nitrogen air gun. This resulted in a rectangular region on one side of the chip thinned down to approximately $\rm20\,\mu m$ and perforated with a series of holes with diameters ranging from $\rm10$ to $\rm50\,\mu m$ (\fig{sample1}a). 

\subsection{CNT growth}

\begin{figure*}[hbtp]
\includegraphics{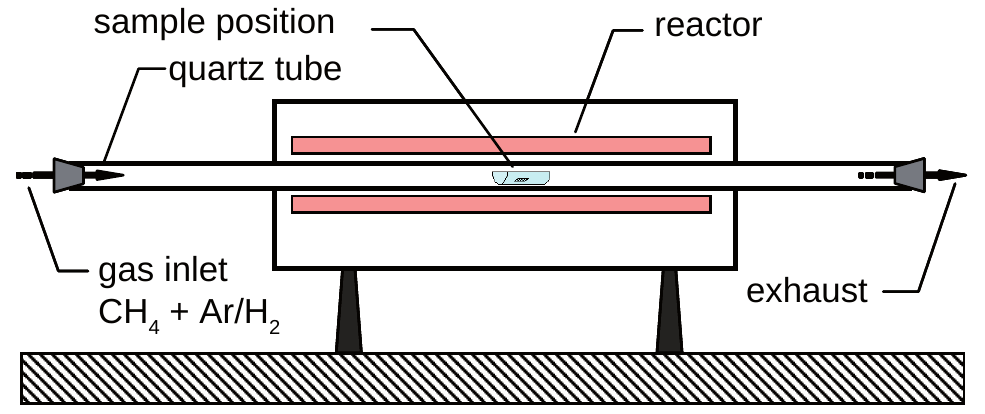}%
\caption{\label{reactor} Sketch of the CVD reactor used for CNT growth.}%
\end{figure*}

\begin{figure*}[hbtp]
\includegraphics{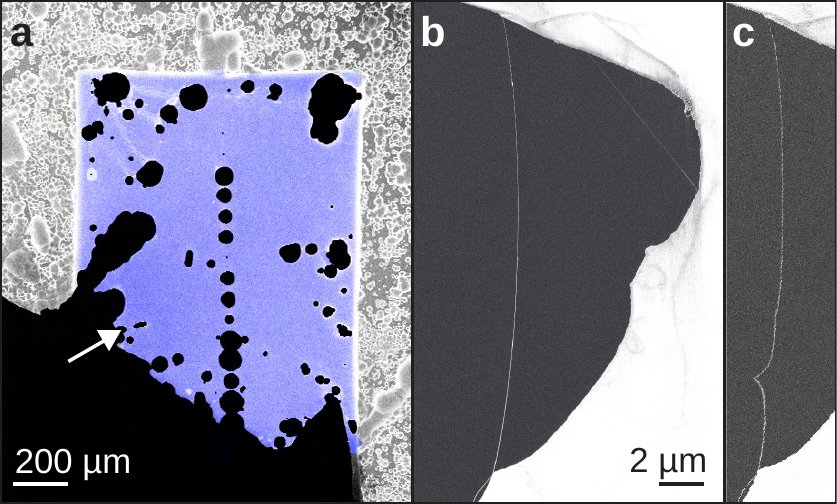}%
\caption{\label{sample1} SEM images of substrate chip and of CNT under investigation. a) Image shows the silicon substrate chip (spotted grey) with thin etched and perforated region (homogeneous, blue) and holes (black). The arrow points at the location of the CNT. b) Zoom showing the CNT under investigation (vertical white line) which is doubly clamped, slightly bent and freely suspended across a distance of $\rm19.7\,\mu m$ in the corner of the substrate (white). c) Kink in CNT occurred once during scanning indicating slack-induced movement and/or the occurrence of conformational changes of the CNT.}%
\end{figure*}

The subsequent CNT growth process was adapted from the method developed by the Schönenberger group\cite{babic_nanolett2003_supp}. The substrate chips were coated with a $\rm170\,nm$ $\rm SiO_2$ layer deposited by PECVD (\fig{samplefab} step 15). In the following, nanocrystalline catalyst particles, a mixture of $\rm Fe(NO_3)_3 - 9H_2O$, $\rm Al_2O_3$ and $\rm MoO_2Cl_2$ (\textit{Sigma-Aldrich}, Iron(III) nitrate nonahydrate, product number: 254223, Aluminum oxide, product number: 551643, Molybdenum(VI) dichloride dioxide, product number: 373710) suspended in isopropanol, were deposited on top of the $\rm SiO_2$ by evaporation of the solvent (step 16). CNTs were grown (step 17) using a CVD process in a quartz tube reactor at $\rm900\,^{\circ}C$ for $\rm10\,min$ at atmospheric pressure (\fig{reactor}). The process gas was a mixture of $\rm Ar/H_2$($\rm 5\,\%$) and $\rm CH_4$ at flow rates of $\rm1.0\,SML$ and $\rm0.5\,SLM$, respectively.

\subsection{SEM/TEM analysis}

\begin{figure*}[hbtp]
\includegraphics{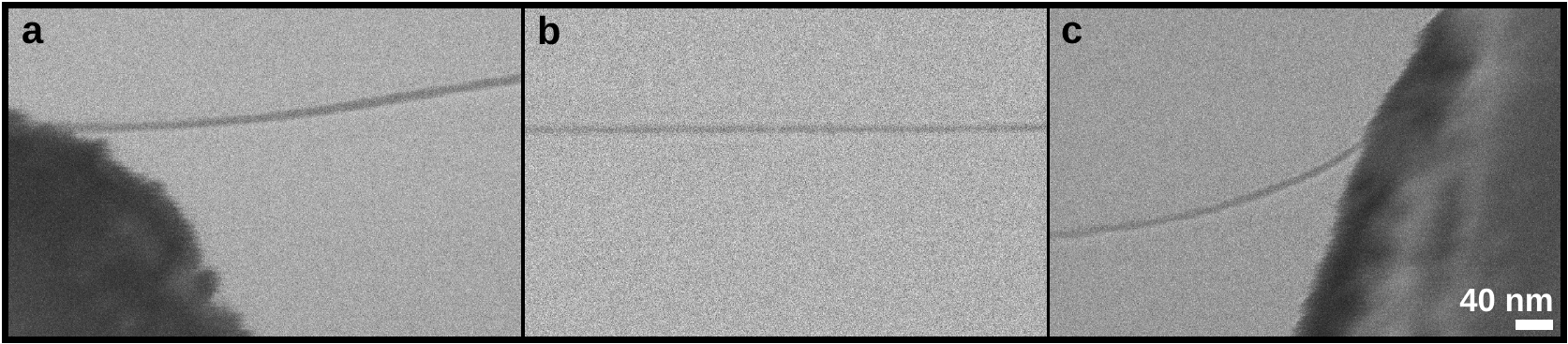}%
\caption{\label{cnt_dimensions} High-resolution SEM images of the CNT discussed in the article at both clamping points (a) and (c) and in the middle (b).}%
\end{figure*}

Careful measurements in the SEM and TEM on different samples show that CNTs grown across gaps as described can have suspended lengths of up to $\rm25\,\mu m$. Diameters of tubes suspended on a length larger than $\rm10\,\mu m$ range from $\rm4$ to $\rm8\,nm$. 
\fig{sample1}b shows an SEM image of the CNT under investigation in this article. The tube's slight wavy shape indicated the absence of tensile stress along the growth axis. Note that this tube moved a little bit during SEM scanning and a conformational change, visible as a kink, occurred and disappeared again (\fig{sample1}c). The tube's diameter was measured from high-resolution SEM images (\fig{cnt_dimensions}) at three positions. It was $\rm8\,nm$ and $\rm7\,nm$ near the clamping points a and c and $\rm6\,nm$ at a central position b, which is consistent with TEM measurements of comparable samples.

\begin{figure*}[hbtp]
\includegraphics{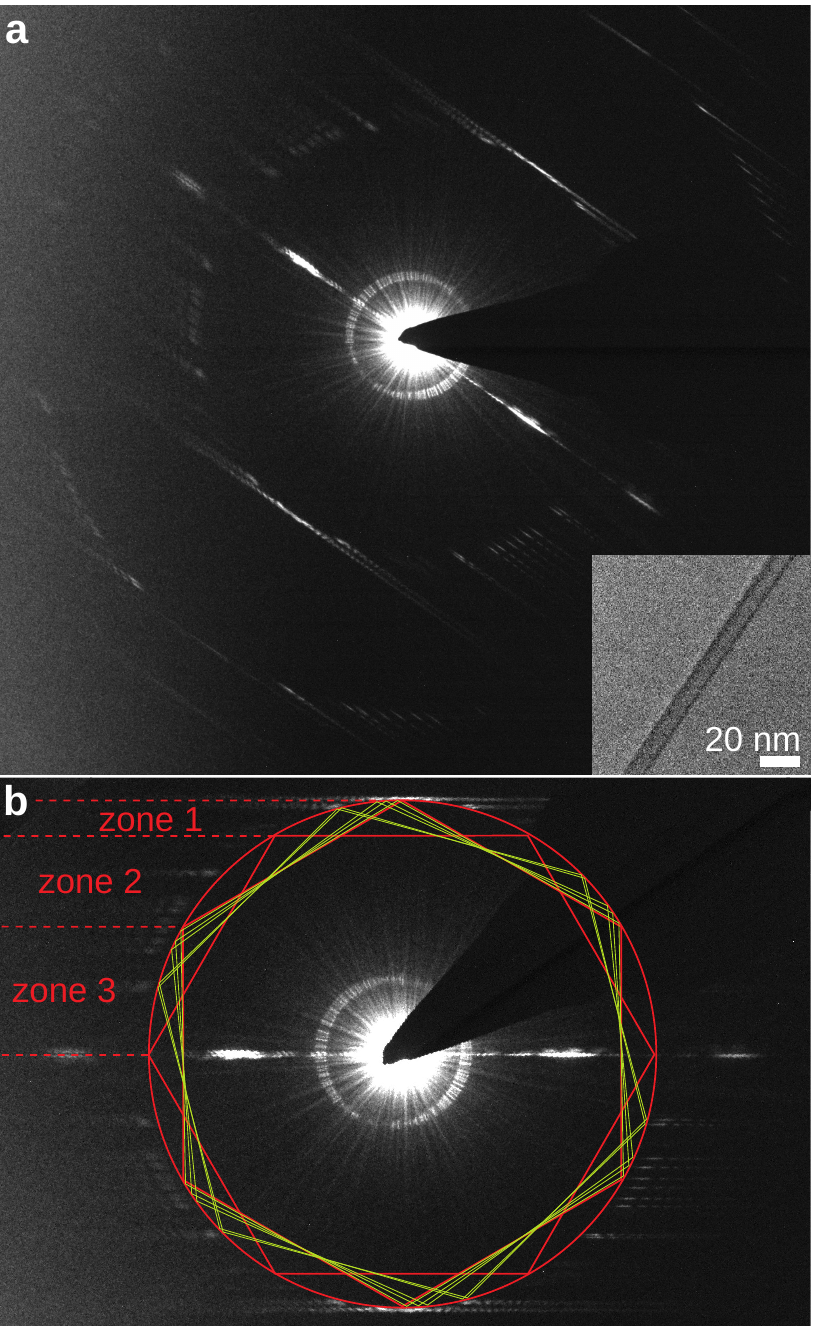}%
\caption{\label{tem-analysis} (a) Electron diffraction pattern of a CNT obtained from TEM measurements. Images of the CNT at the position of the diffraction measurement is shown in the inset. (b) Same diffraction pattern but rotated and with overlays as guide to the eye: Red circle indicates the location of Bragg reflexes. Corners of red hexagons mark points where signatures of CNTs with armchair or zig-zag chirality would occur. These points are used to divide each quadrant of the pattern into three zones. Corners of green hexagons follow Bragg features originating from shells and/or tubes with different chirality composing the CNT. The investigated CNT is thus made up of 7 SWNTs.}%
\end{figure*}

Electron diffraction analysis in TEM makes clear that such long CNTs are not single-walled (SWNT) but rather multi-walled or ropes consisting of $\rm5$ to $\rm7$ individual shells or tubes. An image of a diffraction pattern is depicted in \fig{tem-analysis}a, along with an image of the corresponding CNT in the inset. The diffraction image exhibits horizontally extended Bragg reflexes starting at points on a circle concentrically around the bright central peak (see same diffraction pattern but rotated and with red circle as guide to the eye in \fig{tem-analysis}b). Apart from the black shadow protruding towards the pattern center, which is caused by the beam block, the image is symmetric about the horizontal and vertical axis. We use a simplified version of a  procedure to interpret the diffraction patterns\cite{deniz_ultramic2010}. Therefore each quadrant of the pattern is structured into three zones whose four borders are given by the corners of the red hexagons. These corners indicate the starting points of horizontal Bragg reflexes that would occur if an armchair and a zig-zag SWNT are involved. Each SWNT will exhibit the same diffraction pattern, rotated about an angle depending on chirality. The number of horizontal Bragg reflexes within one zone including its borders is equal to the number of SWNTs contributing to the pattern (see green hexagons in \fig{tem-analysis}b). Thus 7 SWNTs contribute to the displayed diffraction pattern.

\section{Experimental methods and further data}

\subsection{Inserting the CNT chip into the optical cavity}

\begin{figure*}[htbp]
\includegraphics{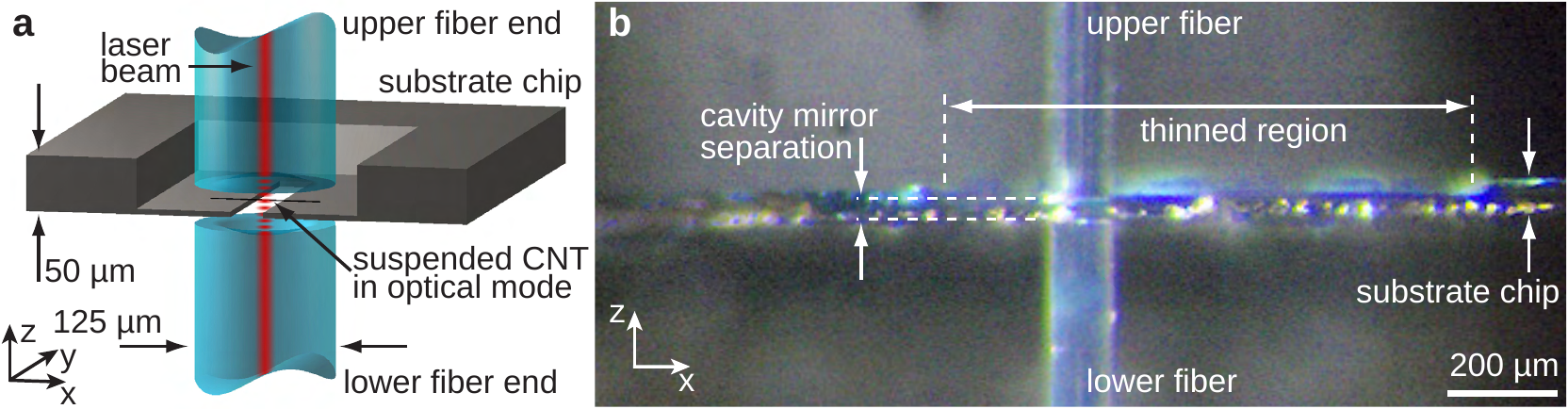}%
\caption{\label{cavityphoto} Schematic (a) and photograph (b) of the cavity with the sample inside. The optical fiber axes are oriented vertically. The sample chip is seen from the side and laying horizontally between the fiber ends forming the cavity mirrors. Regions on the chip appearing much thicker than the cavity mirror separation originate from the far and thus defocussed end of the (slightly tilted) substrate.}%
\end{figure*}

One of the technical key challenges of the experiment is to place the sample into the cavity. Dimensions of the cavity and substrate chip do not leave much room to navigate (compare \fig{cavityphoto}a and b), as the $\rm25\,\mu m$ thick substrate has to be positioned accurately within a $\rm27\,\mu m$ wide cavity gap without touching the fiber mirrors. Technical tolerances from fabrication of the cavity and substrate chip even reduce the clearance. Therefore optical inspection during insertion and positioning of the chip is absolutely necessary. However, the free space between the sample and the fiber ends cannot be resolved in the microscope, as shown in \fig{cavityphoto}b. Looking at the cavity under different viewing angles helps to gain an impression about the actual clearance and thus avoids crashing into the fiber ends.

\subsection{Further spectral data}

\begin{figure*}[htbp]
\includegraphics{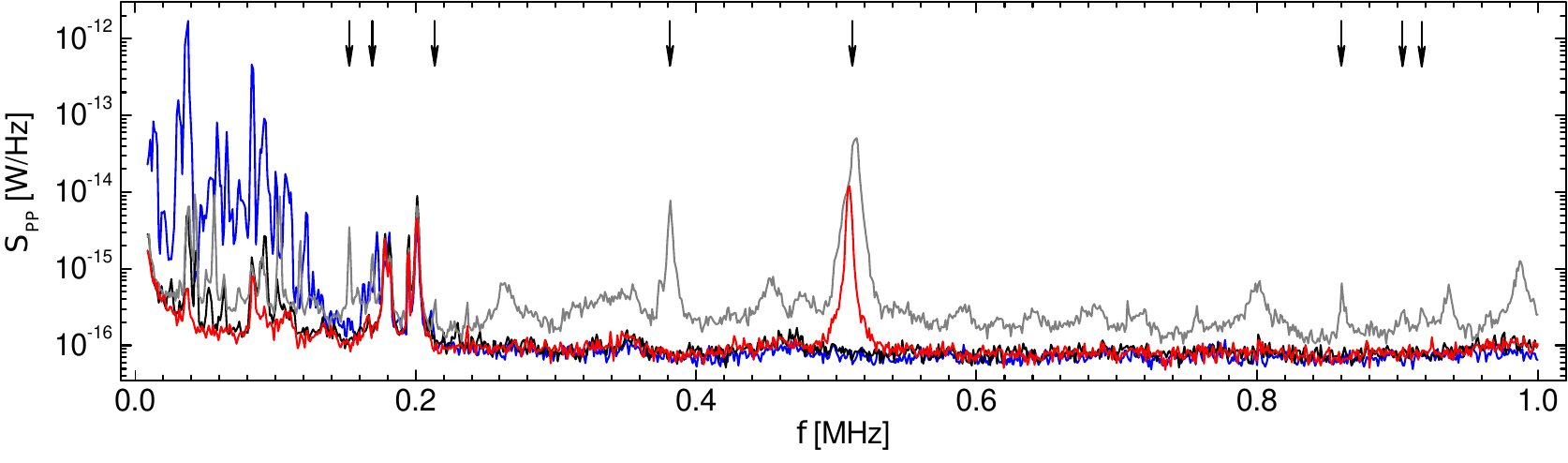}%
\caption{\label{dataplots_supp} Full frequency range of spectra shown in main article (gray and red) with background spectrum (black) and spectrum containing substrate modes (blue). Power spectra reveal the CNT's vibrational spectrum. Background spectra with $\rm14\,dBm$ power of white noise applied to the drive piezo are shown in black with no sample inside the cavity and in blue with the substrate corner (indicated by the black arrow in Fig. 1b of the main article) protruding into the optical mode.  The gray trace is obtained with the CNT placed in the cavity mode and $\rm14\,dBm$ of white noise is applied to the drive piezo, whereas for the red trace the drive is switched off. The remaining peak at about $\rm500\,kHz$ of more than $\rm20\,dB$ above the noise floor is an evidence of the Brownian motion of the CNT. Black arrows point at the peaks identified as mechanical resonances of the CNT in the SEM later. Unmarked peaks below $\rm220\,kHz$ are attributed to substrate modes and technical noise\cite{stapfner_spie2010_supp}.}
\end{figure*}

Under the experimental conditions described in the main text, the optical power transmitted on the empty cavity resonance is $\rm0.56\,\mu W$. Inserting the CNT into the optical mode, as illustrated by the red spot in Fig. 1b of the main article, leads to a $\rm29\,\%$ drop in the cavity transmission. This drop is predominantly caused by residual clipping losses induced by the vicinity of the hosting substrate edges near the relatively short CNT. When locked on the cavity resonance slope, we actuate the drive piezo underneath the sample with white noise ($\rm14\,dBm$ with $\rm20\,MHz$ bandwidth) from a signal generator, and observe a series of resonances appearing clearly above the noise floor (\fig{dataplots_supp} gray trace). Most of the features in the spectrum below $\rm0.22\,MHz$ stem from systematic noise of the laser and cavity stabilization electronics (see Ref. \onlinecite{stapfner_spie2010_supp} for a detailed discussion), and are also present in reference measurements with an empty cavity (black trace). The noise floor above $\rm0.22\,MHz$ limits the sensitivity and has systematic origins such as photodetector noise. At frequencies higher than $\rm0.22\,MHz$, the spectrum shows a series of distinct resonances which do not appear in the empty cavity trace and which can be ascribed to mechanical resonances of the CNT (black arrows). Further resonances which are apparent in the white-noise driven spectrum are most likely of mechanical origin, too, but could not be identified in independent SEM investigations of the same CNT. More specifically, the resonance at $\rm0.51\,MHz$ is the first flexural mode of the tube. This main peak rises up from the noise floor by more than $\rm20\,dB$, even after the drive is switched off (\fig{dataplots_supp} red data). Interestingly, this signal peak is not stable in frequency but fluctuates slowly between $\rm0.47\,MHz$ and $\rm0.52\,MHz$ on the timescale of minutes. We consider slack and conformational instability of the tube, which is also observed in the SEM, to contribute most to this behavior. Furthermore, a strong dependence of the signal on the position of the CNT in the optical mode is observed. It essentially vanishes when the CNT is positioned a few micrometers away from the optical mode axis. Further information can be gained when the sample is positioned such that the optical mode lies on a protrusion of the hosting substrate (for example at the position indicated by a black arrow in Fig. 1b of the main article, about ten micrometers away from the CNT). At this position, with the protrusion entering the optical mode, the drop in cavity transmission is still of order $\rm29\,\%$, but the spectrum (\fig{dataplots_supp} blue trace) exhibits extra peaks below $\rm0.22\,MHz$, most probably orginating from vibrating substrate modes. Since none of these substrate resonances are observed above $\rm0.22\,MHz$, it can be concluded that the main resonance at $\rm0.51\,MHz$ is not associated with a susbstrate mode. Even though tight geometric constraints (see \fig{cavityphoto}b) prevent us from performing a complete position scan of the CNT along the z-axis of the cavity revealing the half-wavelength periodicity of the Fabry-Pérot cavity mode, we observe a constant value for the cavity resonant transmission at several z-positions. This indicates that the presence of the CNT is not the dominant contribution to the total extra cavity losses in our experiments but rather clipping losses of the neighboring substrate edges. Thus the dispersive cavity response rather than dissipative interaction is crucial for displacement detection of the CNTs vibrational modes. To extract information on the nanomechanical system from the dissipative signal component, devices with larger gaps need to be engineerd.

\section{Effective mass}

The mathematical construct of the effective mass aims at reducing the full differential equation of an elastic beam to the simple equation of motion of a harmonic oscillator. While elastic beam theory results in the time-dependent deflection of each point along the beam, the harmonic oscillator simplifies the calculation to the case of a point-mass oscillating at a specific location on the beam. The effective mass thus describes the fraction of the beam's total mass corresponding to the amplitude of the harmonic oscillator which is equivalent to the deflection of the beam. This solution is more practical to work with, but requires some preliminary work. As will be made clear, the effective mass depends on the reference position along the beam, the geometry of the deflection measurement device and the distribution of the driving force and thus may depend on the selected driving and detection scheme. Calculations in this section follow calculations presented in Ref. \onlinecite{timoshenko_supp, cleland, Karabacak_PHD, Anetsberger_PHD}.

\subsection{Formalism}

Starting from the Euler-Bernoulli equation of an elastic flexural beam with neither tensile nor compressive stress, which is driven by a time and position dependent force $f(x,t)$

\begin{equation}
  \frac{\partial^2}{\partial x^2}E\,I \frac{\partial^2 U(x,t)}{\partial x^2} + \rho\,A\frac{\partial^2 U(x,t)}{\partial t^2} = f(x,t)
\label{euler-bernoulli}
\end{equation}

the time and position dependent deflection reads

\begin{equation}
  U(x,t)=\sum_n q_n(t)\,\phi_n(y).
\label{deflection}
\end{equation}

$U(x,t)$ superimposes the product of the time dependent amplitude $q_n(t)$ and the position dependent mode shape function $\phi_n(x)$ for all modes $n$. Further parameters in \eq{euler-bernoulli} are the Young's module $E$, the area moment of inertia $I$, the mass density $\rho$ and the cross-section area $A$. Multiplying \eq{euler-bernoulli} by the shape function $\phi_m(x)$ for mode $m$, integrating over the beam length $l$ and exploiting the orthogonality of the shape functions

\begin{equation}
 \int_0^l\phi_m(x)\,\phi_n(x)\,dx = 0 \quad \mathrm for \quad m\neq n
 \label{orthogonality}
\end{equation}

as well as the identity relation

\begin{equation}
  \int_0^l\phi_n(x)\frac{\partial^4\phi_n(x)}{\partial x^4}\,dx=\int_0^l\left(\frac{\partial^2\phi_n(x)}{\partial x^2}\right)^2\,dx
\label{identity_rel}
\end{equation}

yields

\begin{equation}
  E\,I\,q_n(t) \int_0^l\frac{\partial^2\phi_n(x)}{\partial x^2}\,dx + \rho\,A\,\ddot{q}_n(t) \int_0^l\phi_n^2(x)\,dx = \int_0^l f(x,t)\,\phi_n(x)\,dx.
\label{}
\end{equation}

This looks like a harmonic oscillator

\begin{equation}
  m_n^\prime\,\ddot{q}_n(t) + k_n^\prime\,q_n(t) = \int_0^l f(x,t)\,\phi_n(x)\,dx
\label{oscillator}
\end{equation}

with mass and spring constant

\begin{subequations}
\begin{align}
  m_n^\prime &= \rho\,A \int_0^l\phi_n^2(x)\,dx  = m \int_0^1\phi_n^2(\xi)\,d\xi,
\label{mn}\\
  k_n^\prime &= E\,I \int_0^l\frac{\partial^2\phi_n(x)}{\partial x^2}\,dx = \frac{E\,I}{l^3} \int_0^1\left[\frac{\partial^2\phi_n(\xi)}{\partial\xi^2}\right]^2\,d\xi,
\label{kn}
\end{align}
\end{subequations}

normalized coordinate $\xi = x/l$ and the mode function $\phi_n(\xi))$ normalized to $1$ on the interval $\xi\in[0;1]$.

In most physical deflection measurement devices the detection is not carried out at a single point but rather distributed over a certain region along the beam. To account for that the mode shape function $\phi(\xi)$ is convoluted with a distribution $\eta^2(\xi)$. Thus the measured deflection $u_n(t)$ reads

\begin{equation}
  u_n(t) = q_n(t)\int_0^1\phi_n(\xi)\eta^2(\xi)\,d\xi,
\label{integrated_deflection}
\end{equation}

where the distribution function $\eta^2(\xi)$ obeys the normalization

\begin{equation}
  \int_{-\infty}^{\infty}\eta^2(\xi)\,d\xi=1. 
\label{normalization}
\end{equation}

Multiplying \eq{oscillator} with the convolution integral on the right hand side of \eq{integrated_deflection} gives

\begin{equation}
  m_n^\prime\,\ddot{u}_n(t) + k_n^\prime\,u_n(t) = \int_0^l f(x,t)\,\phi_n(x)\,dx\,\int_0^1\phi_n(\xi)\eta^2(\xi)\,d\xi. 
\label{}
\end{equation}

Assuming a driving force equally distributed along the beam $F(t) = \bar{f}(t)\,l$ such that $f(x,t)\mapsto\bar{f}(t)=F(t)/l$ does no longer depend on $x$ and $\int_0^l f(x,t)\,\phi_n(x)\,dx = F(t)\int_0^1 \phi_n(\xi)\,d\xi$ for normal coordinate $\xi$, the effective harmonic oscillator becomes

\begin{equation}
  m_{\text{eff},n}\,\ddot{u}_n(t) + k_{\text{eff},n}\,u_n(t) = F(t).
\label{eff_oscillator}
\end{equation}

The effective mass $m_{\text{eff},n}$ and the effective spring constant $k_{\text{eff},n}$,

\begin{subequations}
\begin{align}
  m_{\text{eff},n} &= m\frac{\int_0^1(\phi_n(\xi))^2\,d\xi}{\int_0^1 \,\phi_n(\xi)\,d\xi\,\int_0^1\phi_n(\xi)\eta^2(\xi)\,d\xi} \quad \mathrm and
  \label{meff}\\
  k_{\text{eff},n} &= \frac{E\,I}{l^3}\frac{\int_0^1\left[\frac{\partial^2\phi_n(\xi)}{\partial\xi^2}\right]^2\,d\xi}{\int_0^1 \,\phi_n(\xi)\,d\xi\,\int_0^1\phi_n(\xi)\eta^2(\xi)\,d\xi},
  \label{keff}
\end{align}
\end{subequations}

comprise the inertial and elastic properties of the whole beam oscillating in mode $n$. This allows modeling of the beam's dynamics with a simple harmonic oscillator. A valid calibration of the measured amplitude is possible with the knowledge of the measurement distribution.

\subsection{Application to the experiment}

In the following, the discussed formalism of the effective mass is applied to the situation presented in the main article. The CNT is modeled as a doubly-clamped beam with rigid clamps and without tensile or compressive stress. Boundary conditions for this case are 

\begin{equation}
  \phi_n(0)=\phi_n(l)=\left.d\phi_n(x)/dx\right|_{x=0}=\left.d\phi_n(x)/dx\right|_{x=l}=0,
\label{boundary_cond}
\end{equation}

implying the deflection and slope of the CNT are both zero at the clamps. This results in a mode shape function

\begin{equation}
  \phi_n(x)=a_n \left(\cos(\beta_n x)-\cosh(\beta_n x)\right)+\left(b_n \sin(\beta_n x)-\sinh(\beta_n x)\right)
\label{mode_shape_fct}
\end{equation}

with mode shape roots $\beta_n l$ and coefficient ratios $a_n/b_n$ as given in \tab{coeffs} for the first five modes\cite{timoshenko_supp}. The coefficient $a_n=1$ is chosen for all $n$ and $\phi(x)$ is normalized to $1$ on the interval $\xi\in[0;1]$.

\begin{table}[htbp]
		\begin{tabular}{|l|r|r|r|r|r|}
			\hline
			mode index, $n$ 					  	&	0				&	1				&	2				&	3				& 4	\\
			\hline
			mode shape roots, $\beta_n l$	&  4.7300 &  7.8532 & 10.9956 & 14.1372 & 17.2788 \\
			\hline
			coefficient ratio, $a_n/b_n$	& -1.0178 & -0.9992 & -1.0000 & -1.0000 & -1.0000 \\
			\hline
		\end{tabular}
		\caption{Numerical values of $\beta_n l$ and $a_n/b_n$ for the first five modes $n$ of the doubly rigidly clamped beam.}
		\label{coeffs}
\end{table}

The CNT's deflection is probed by the optical cavity mode, which is a laser field with Gaussian distribution and waist radius $w_0$. Thus 

\begin{equation}
  \eta^2(x)=\frac{1}{2\,w_o\sqrt{2\,\pi}}\,\mathrm e^{\frac{-2(x-x_0)^2}{w_0^2}}
\label{cavity_mode}
\end{equation}

with normalization \eq{normalization}. 

The Brownian motion of the CNT is driven by a thermally induced force distributed equally along the tube stipulating $f(x,t)=F(t)/l$ and no longer depends on the $x$ coordinate.

Entering $\phi_0(\xi)$ and $\eta(\xi)$ in normalized coordinates $\xi=x/l$ and constants for the parameters specified in the main article $l=19.7\,\rm\mu m$ and $w_0=3.4\,\rm\mu m$ allows computation of the effective mass of the employed CNT. For the first flexural mode $n=0$ it is found to be $m_{\text{eff}}= 0.81\cdot m$ with the total mass $m$ of the CNT.


\end{document}